# Tracing the dynamics of superconducting order via transient third harmonic generation

Min-Jae Kim[1,2,3], Sergey Kovalev[4], Mattia Udina[5], Rafael Haenel[2,6], Gideok Kim[2], Matteo Puviani[2], Georg Cristiani[2], Igor Ilyakov[4], Thales V. A. G. de Oliveira[4], Alexey Ponomaryov[4], Jan-Christoph Deinert[4], Gennady Logvenov[2], Bernhard Keimer[2], Dirk Manske[2], Lara Benfatto[5] and Stefan Kaiser[1,2,3]

[1] Institute of Solid State and Materials Physics, Technical University Dresden, 01062 Dresden, Germany
[2] Max Planck Institute for Solid State Research, 70569 Stuttgart, Germany
[3] 4th Physics Institute and Research Center SCoPE, University of Stuttgart, 70569 Stuttgart, Germany
[4] Helmholtz-Zentrum Dresden-Rossendorf, 01328 Dresden, Germany
[5] Department of Physics and ISC-CNR, "Sapienza" University of Rome, 00185 Rome, Italy
[6] Department of Physics and Astronomy & Stewart Blusson Quantum Matter Institute, University of British Columbia, Vancouver BC V6T 1Z4, Canada

*email: mj.kim@fkf.mpg.de, lara.benfatto@roma1.infn.it, stefan.kaiser@tu-dresden.de

**Ultrafast optical control of quantum systems is an emerging field of physics. In particular, the possibility of light-driven superconductivity with ultrashort laser pulses has attracted much of attention. To identify non-equilibrium superconductivity, it is necessary to measure fingerprints of superconductivity on ultrafast timescales. Recently non-linear THz third harmonic generation (THG) was shown to directly probe the collective degrees of freedoms of the superconducting condensate including particularly the Higgs mode. Here we extend this idea to light-driven non-equilibrium states in superconducting $La_{2-x}Sr_xCuO_4$ establishing a protocol to access the transient superconducting (SC) order-parameter fluctuations. We perform an optical pump-THz-THG drive experiment and use a two-dimensional spectroscopy approach to disentangle the driven third-harmonic response of optically excited quasiparticles and the pure condensate response. In this way, 2D spectroscopy separately probes both the ultrafast pair breaking dynamics and transient pairing amplitude of the condensate.**

The ability of short light pulses to stabilize quasi-ordered states not accessible in thermal equilibrium demonstrated its potential in a variety of fields. Resonant ultrafast excitation of atomic oscillations controls the electronic properties of materials, leading to remarkable phenomena such as switching of ferroelectricity and multiferroicity [1-4] and light-induced superconductivity [5-8]. In the case of ordered states formed via the spontaneous breaking of a pre-existing symmetry, time-resolved protocols turned out to be crucial to address the physical mechanisms at play in unconventional systems, which deviate from the standard mean-field like description. A paradigmatic case is provided by high-temperature superconductors like cuprates, where superconductivity emerges out of a complex and strongly correlated normal state with marked signatures of fluctuating charge and spin order [9]. In this situation, standard spectroscopic probes can fail in providing a clear-cut signature of the pure superconducting order, and of its dynamical evolution. At the same time, time-resolved protocols in both conventional and unconventional superconductors revealed so far rather different dynamics depending on the wavelength of the initial pump pulse. Indeed, the ability of a short light pulse to quench the superconducting order is partly hindered by the simultaneous creation of quasiparticle excitations, leading to a mixing among the two signals that can be hard to disentangle [10-14].

Here we outline a new approach to pinpoint the dynamics of the superconducting order parameter via transient detection of THz driven third-harmonic generation (THG). In standard THG measurements, a high-field multicycle THz pulse of frequency $\omega$ below the superconducting gap enforces driven oscillations of the SC condensate with twice the driving frequency, leading to a characteristic third harmonic (TH) generation signal at frequency $3\omega$. As shown by previous work, the THG signal can be well accounted by a quasi-equilibrium approach, where the strength of the $3\omega$ response scales with a non-linear optical kernel $\chi$ computed *at equilibrium* in the condensate. In the SC state $\chi$ is dominated by Cooper-pair and Higgs mode excitations at $2\Delta$, i.e. twice the equilibrium value of the SC order parameter [15-24]. In the present work, we use a second intense optical pump to modulate the SC ground state and then dynamically access the transient evolution of the THG. We show that by means of the simultaneous control over the gate acquisition time and the pump-probe delay we can efficiently disentangle the quasiparticle pair-breaking effects from the pure time evolution of $\Delta(t)$. Further, realizing a well understood quench protocol of the condensate we probe its recovery on a picosecond (ps) timescale. Our experiment demonstrates the high potential of two-dimensional optical-THz pump-drive experiments in order to disentangle the various degrees of freedom at play in a photoexcited process, via a selective identification of the relevant excitation mechanisms.

**Results**

In this study we report the light induced non-linear THz response of photoexcited La$_{2-x}$Sr$_x$CuO$_4$, a high-$T_c$ superconducting cuprate. We extend the THz driven TH generation experiments [16, 25] into a pump-drive experiment in the time domain as sketched in Fig. 1a. As done in previous studies [25] we initially generate TH without optical excitation shown in as an electro-optical (EO) sampling trace with the internal gate delay time $t_g$ in Fig. 1b: A high-field multicycle THz drive $E_{THz}(\Omega)$ at ω=0.7 THz (2.5 meV) far below the superconducting gap as driving pulse generates a TH field $E_{THz}(3\Omega)$ at ω=2.1THz in the superconducting La$_{2-x}$Sr$_x$CuO$_4$ thin film at 12K ($T_c$=44K, Δ$_{SC}$~20 meV [26]). As previous experimental [16, 17, 25, 27-31] and theoretical work [15-24, 32, 33] demonstrated, the thermally induced transition to the SC state strongly enhances the THG. Within a quasi-equilibrium approach the THG is controlled by the non-linear current j$_{NL}$~$\chi$E$^3$, where $\chi$ is the non-linear kernel and E is the local electric field within the sample. In the presence of a long-range SC order parameter $\Delta$ the non-linear response $\chi$ shows a marked resonance at 2$\Delta$ due to both particle-hole BCS excitations and Higgs oscillations on top of the equilibrium SC ground state. Even though their relative importance depends on various material parameters, like e.g. disorder [19, 20, 24] and correlations [22], and they can coexist with additional low-energy SC fluctuations as e.g. Leggett [29, 32] and soft plasma modes [34], it is well established that in single-band superconductors the THG enhancement below $T_c$ follows the rising of $\Delta$. In our experiment a femtosecond optical pump pulse at 1.55 eV excites the superconductor and the light-induced changes of the TH signal probe the triggered non-equilibrium dynamics of the condensate. The transient dynamics is scanned by varying the time delay $\Delta t$ between the pump and drive pulses. Two experimental protocols are possible. In a 2D scan, full TH transients are measured for different pump delays. In a one-dimensional (1D) scan, only the pump beam is moved with respect to the TH field that is probed at a specific position. For the experimental details, see Supplementary Note 1.

Figure 1c shows the photoexcited TH signal dynamics measured as a 1D scan taken as changes in the peak amplitude of the unperturbed TH field $E_{THz}(3\Omega)$ (marked by the arrow in Fig. 1b) in the superconducting state below $T_c$. Varying the pump-drive delay $\Delta t$ the TH sampling position remains fixed. The differential TH signal $\delta E_{TH}$ is characterized by an *increase* of the THG, with a long-lasting incoherent excitation dynamics and an additional oscillatory signal at twice ($2\Omega$) and fourth ($4\Omega$) the frequency of the driving field ($\Omega$) at early pump-drive time delay (Supplementary Note 2). In the following such higher harmonics will be characterized as sidebands at ($3\Omega \pm \Omega$) of the original TH signal due to a wave mixing with the visible pump pulse. This is done via a 2D spectroscopic approach measuring the generated transient TH at different pump-drive delays $\Delta t$ (Fig.1a).

The incoherent dynamics is characterized by a rise-time of ~5ps and a fast decay time of about 20 ps by a bi-exponential fit for a low pump excitation fluence of 15µJ/cm². These time scales are more than twice longer than the known quasiparticle relaxation time scales under photoexcitation in a linear THz or optical response in LSCO and YBa$_2$Cu$_3$O$_{7-\delta}$ (YBCO) [10, 14]. Also trARPES (time- and angle- resolved photoemission spectroscopy) shows an excitation and relaxation dynamics of the gap in the low ps range for Bi$_2$Sr$_2$CaCu$_2$O$_{8+\delta}$ (BSCCO) [35]. The increase of the TH signal is at first sight rather surprising: indeed, THG measurements at equilibrium demonstrated that the non-linear kernel $\chi$ increases in the SC state, suggesting that a partial melting of SC order due to the pump should rather lead to a suppression of THG. However, as mentioned above, the measured TH amplitude itself needs to be carefully examined, since the incoherent dynamics of excited quasiparticles can strongly affect the screening of the local field $E$. This effect, rather than a genuine increase of the non-linear kernel $\chi$, could explain the increasing of the TH field, as recently argued in [36]. As a consequence, the extraction of the non-linear response requires a proper normalization of the signal [25], as we will discuss below.

To characterize the spectral components of the differential TH signal, we perform 2D scans to obtain the full 2D spectrum of the transient TH dynamics as a function of the internal gate time of the EO-sampling, $t_g$ and the pump-drive delay time $\Delta t$. Figure 2a shows this spectrum in time-domain for field strength of 100 kV/cm of the multi-cycle THz pulse and a pump fluence of $F_{pump}$=20 $\mu J/cm^2$ of the femtosecond optical pump. The vertical pattern along $t_g$ for a fixed $\Delta t$ shows the response to the THz drive: as in the non-pumped experiments [25], the transmitted THz signal shows oscillations as a function of $t_g$ consisting of a superposition of the fundamental harmonic (FH) and the TH component generated by the non-linear response inside the superconductor. They are both clearly visible in Fig. 2b, where the vertical axis $f_g$ is the Fourier transform of the gate time, and marked bands appear at $f_{FH}$ = 0.7 THz and $f_{TH}$ = 2.1 THz, respectively. When the optical pump sets in, a modulation of the TH along $\Delta t$ appears, persisting for the pulse duration of the drive. The modulation frequency of the TH signal, already seen at early times in the 1D scans in Fig. 1c, has a strong component at $2\Omega$ and a smaller one at $4\Omega$ of the driving pulse (see Supplementary Note 2). In addition, a strong transient signal at a frequency lower than $3\Omega$ around $f_g$ =1.4 THz emerges in Fig. 2b. Figure 2c shows the response along the 2D Fourier transformed coordinates, $f_g$ and $f_{\Delta t}$, respectively. Here we can disentangle the different contributions arising from THz high-harmonic generation, transient incoherent dynamics and THz-optical wave mixing processes. The FH and TH of the driving THz field at $f_{\Delta t} = 0$ are marked by the blue and violet circles. These are the same features present in the non-pumped experiment [25]. The pump-probe incoherent dynamics responsible for the long-lasting increase of δ$E_{TH}$ shown in Fig. 1c contributes to the regions highlighted by the blue and violet cigar shapes. The new harmonical features that we observe, both the TH modulation in Fig. 1c and 2b, as well as the new spectral feature around 1.4THz in Fig. 2b, can be understood as sideband processes in the periodically driven system. Such a response, absent in the transient optical pump THz probe experiment [14], can be understood by considering 4- and 6-wave mixing processes among the THz and optical pulses (see Supplementary Note 3), leading to spectral features at $f_g = f_{\Delta t} \pm \Omega$ (green rectangles) and $f_g = f_{\Delta t} \pm 3\Omega$ (red rectangle). The former, in particular, is responsible for a modulation of the TH signal with a periodicity of $2\Omega$ (and $4\Omega$ at higher order) while the drive pulse is present. The spectral width of the sidebands is mainly set by the width of the difference-frequency process around zero frequency of the femtosecond optical pulse.

Notice that the present second harmonic (SH) modulation of the TH signal, I$_{2w}$, can be still fully captured within a quasi-equilibrium approach, shown in Fig. 2d and it is not linked to any (static of dynamically-generated) additional zero-frequency field component, usually invoked to explain SH generation of the original driving field in non-pumped experiments [30, 37, 38]. In Fig. 2d we show its temperature dependence (green, in log scale), alongside with the static TH field intensity (grey, linear scale, from ref. [25]). Both decrease as the temperature increases and they both are detected even above $T_c$. The nonvanishing TH signal above $T_c$ has been previously reported [25] and may be linked to superconducting fluctuations of various origin [34, 39-41].

Having characterized the full 2D spectrum let us focus on the fluence and temperature dependence of the differential TH signal. Therefore, the pump fluence and temperature dependence of the TH response are investigated by optically filtering only for the TH dynamics in 1D pump scan (Fig. 1b and 1c). Fig. 3a and 3b show the dynamics of the maximum TH field changes δ$E_{TH}$ (sampled as in Fig. 1b) under the photoexcitation as a function of excitation density and temperature, respectively. In order to measure the excitation density dependence, in Fig. 3a we fix the temperature at 12K and we sweep the optical-pump fluence from 15 to 50 µJ/cm$^2$. For the temperature dependence reported in Fig. 3b we set a pump fluence of 20 µJ/cm$^2$ that is high enough to saturate the contribution of excited quasiparticle to the linear THz response ( $F \geq 7$ µJ/cm$^2$ [14]) but lower than the threshold where thermal suppression of the SC condensate is known to occurs [14]. We note that in Ref. [14] the data is interpreted that at this fluence the SC state is completely suppressed. However, here we have evidence of a SC state from the THG signal.

As a function of increasing pump fluence the light-induced changes of TH at long time delays, along with the amplitude of the short-time oscillations, monotonically increase. However, the peak value of the TH-field change $\delta E_{TH}^{MAX}$ shows a non-monotonic behavior as a function of the pump fluence (Details of the fit procedure are explained in Supplementary Note 2). Figure 3c shows that at low fluences $\delta E_{TH}^{MAX}$ increases with the square root of the intensity and therefore with the pump field $E_{pump}$. Above a critical value,

$F_c$=30μJ/cm², $\delta E_{\text{TH}}^{\text{MAX}}$ stays positive, but we observe a linear decrease superimposed to the square-root increase. Such a response is characteristic for a depletion of a condensate as was also observed in a similar scenario of photo-excited excitonic insulators [42, 43]. As mentioned above, the $E_{TH}$ field scales as $E_{TH} \sim j_{\text{NL}} \sim \chi E^3$, where $E$ is the local electric field. The square root increase follows the enhancement of the local field $E$ inside the sample due to the pump field. The linear decrease at $F > F_c$ is due to emergence of additional pair-breaking processes in the condensate, which suppress the intrinsic non-linear kernel $\chi$. This characteristic decrease shows that we directly probe the condensate. In contrast, pump-probe spectroscopies probing the linear response to either a THz or an optical pulse [14] on the contrary always show a monotonic increase, even beyond the critical pump fluence for saturation of the SC component. This is due to the fact that a linear probe is only sensitive to the excited quasiparticle avalanche generated in the pair breaking process [14, 44, 45].

To gain further insight into the simultaneous effects due to the quasiparticle excitations and the order-parameter suppression on the TH we turn to the temperature dependence in Fig. 3b. The extracted differential peak value of the TH signal $\delta E_{\text{TH}}^{\text{MAX}}$ at low fluence is shown as closed circles in Fig. 3d. A positive differential TH signal persists up to a temperature of $T \approx 30$ K, and changes its sign for higher temperatures. On further increasing temperature the negative differential TH response remains small even above $T_c$ before it vanishes.

Previous scaling analysis of equilibrium THz response has shown that in optimally doped LSCO the scaling fluctuation frequency becomes nonzero already below $T_c$, at around 30K [46]. Interestingly such an onset also appears in the temperature-dependent unpumped TH amplitude response, that shows a strong screening peak at 30K [25] (see Supplementary Note 4). This finding suggests that the sign-change of the pump- induced variation of the TH signal $\delta E_{\text{TH}}^{\text{MAX}}$ can be ascribed again to a predominant effect of the local-field screening, such that a dip occurs at the same temperature where a peak is found in the dissipative linear THz response [46]. An average heating effect can be ruled out by the temperature-dependence of $\delta E_{\text{TH}}^{\text{MAX}}$ in the high fluence regime $F > F_c$ shown as open circles in Fig. 3d. It traces the same temperature dependence of the low-fluence data up to 30K, but since it is ultimately dominated by SC condensate suppression it is barely affected by the local-field screening effects above 30K, so that the $\delta E_{TH}^{MAX}$ signal remains zero above it.

Now we can investigate the intrinsic SC order-parameter response under photoexcitation with the low pump fluence of 20 µJ/cm$^2$ at 5K. In the 2D protocol used for Fig. 2a we keep the distance $\Delta t$ between the center of the THz pulse and the probe pulse fixed while scanning on the acquisition time $t_g$. Here we keep the distance $\tau$ between the acquisition time and the pump pulse fixed, as e.g. done for transient THz spectroscopy [5, 7, 47] and described e.g. theoretically in ref. [48] (see Supplementary Note 5). Therefore, since $\tau = t_g + \Delta t$ in the 2D measurement of Fig. 2a we shear the data in the x-axis ($\Delta t$), as shown in Fig. 4a, in order to obtain a 2D scan as a function of $t_g$ vs $\tau$. Then we perform the Fourier-transformation along the gate time $t_g$ to reveal the transient spectrum in Fig. 4b. Here, the FH component now shows strong modulations as a function of pump-drive delay time $\tau$, while the modulations in the raw TH signal remain weak. The FH modulations show broad frequency contributions around 0.7 THz to 1.4 THz in a 2D Fourier transform spectrum (see Supplementary Note 5). This broad feature is intrinsically linked to the 45 degree tilted wave mixing signal in Fig. 2c. The full 2D transient FH and TH responses show the direct observation of the non-linear current generation by excited quasiparticles with the driving THz field. Now, with the simultaneous determination of FH and TH we can estimate the pump effects on the non-linear optical kernel $\chi$ by normalizing the TH field amplitude to the transmitted FH field $I_{TH} / I^\beta_{FH}$ [25, 31, 37]. As shown in Fig. 4c the envelop of the resulting TH susceptibility $\chi \sim (I_{TH}(\tau))/(I_{TH}^{eq})$ shows now a marked suppression at short time delays, along with modulations originating from the FH signal in Fig. 4b. Once eliminated the screening effects on the local electric field, $\chi$ traces the transient SC order-parameter response, with a full suppression of the signal within 1ps and a relaxation back to the equilibrium value in about 9ps. These time scales are well in agreement with the fast picosecond suppression and recovery of the SC condensate under photo-excitation probed by trARPES for other cuprate superconductors [49]. Even though we cannot resolve the different contributions of amplitude and phase fluctuations of the SC order parameter made possible by trARPES, we are nonetheless able to show that 2D transient spectra directly probe the fast dynamics of the condensate, without spurious effects due to quasiparticle-excitations processes present in 1D scans.

**Discussion**

Probing the transient non-linear THG response of light driven LSCO provides novel insight into the multiple contributions of the superconducting non-linearities and pair breaking processes to the THG signal. In static THG measurements these are difficult, and often cannot be unambiguously identified and disentangled [15-24] Here, transient THG changes under photo-excitation in an optical pump - THz drive scheme allow us to separate the responses of quasiparticles and the SC condensate in cuprate superconductors under optical photo-excitation.

Modulations of the signal at $2\Omega$ and $4\Omega$ are fully understood as sideband modulation of the signal due to four-wave mixing among the optical pump and the THz probe field. The excitation density- and temperature- dependent measurements show how the pump-induced quasiparticle excitations modify the screening of the local field, accounting for the THG increase in the low-fluence regime and for its sign-change across the onset temperature for thermal fluctuations, where both the onset temperature and pair-breaking thresholds are in agreement with existing literature. At larger fluence the THG signal becomes suppressed, signaling the suppression of the SC condensate with a reduction of the THG. In contrast to linear spectroscopy, our probing scheme allows us to identify the pure effects ascribed to suppression of SC order, that only occurs above a fluence threshold. The effect becomes clear in the 2D scans, where THG shows a suppression and recovery over a very fast (~9 ps) time scale. As mentioned above, it is well accepted that the non-linear optical kernel in superconductors grows below $T_c$ along with the increase of the SC order parameter Δ. As a consequence, the suppression of the THG signal in 2D scans is likely to be due to a direct suppression of the intrinsic order-parameter amplitude (Higgs) contribution to the non-linear response. Nonetheless, it is still an open question whether the same effect could be ascribed to the collapse of superconductivity due to a quench of the SC phase coherence, as suggested by trARPES data [49]. Even though transient THG cannot easily resolve the amplitude and phase dynamics as it happens via direct measurements of the single-particle spectral function, the acquisition of full 2D scans over an extended temperature range could elucidate this issue. On a similar avenue, 2D THz experiments with broad-band pulses [50] could serve as probes for the transient regime. 2D transient THG spectroscopy thus offers a novel tool to investigate the quenching and recovery of the Higgs and phase excitations of the SC condensate in unconventional superconductors, and shedding light on the fundamental pairing mechanism at play in these systems.

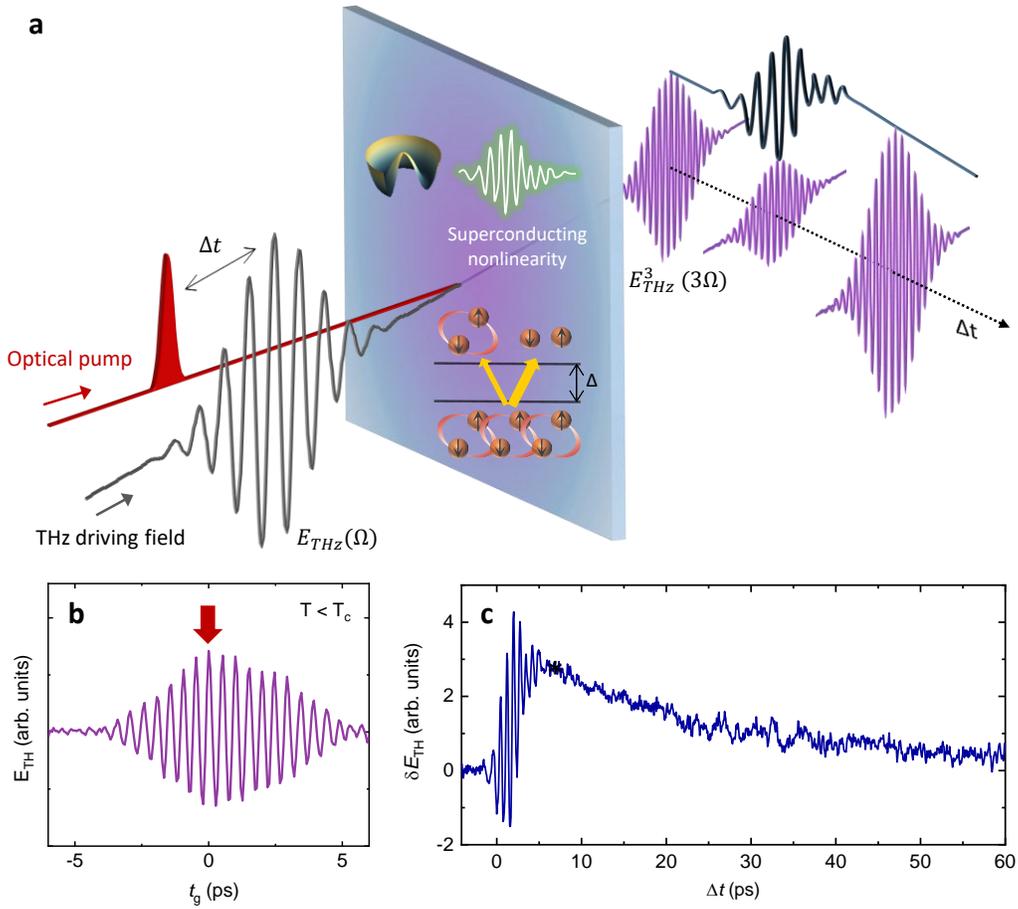

**Figure 1 Optical pump-TH drive signal in a superconducting LSCO thin film.** (a) Experimental scheme: An optical pump pulse triggers the pair breaking dynamics in the superconductor. A time delayed intense multi-cycle THz driving field $E_{THz}(\Omega)$ drives a third harmonic generation $E^3_{THz}(3\Omega)$ that probes the transient non-linear dynamics at a pump time delay of $\Delta t$ after the excitation pulse. The transient dynamics of the TH field results in a superposition of intrinsic THG changes of the superconducting condensate itself and contributions of nonlinear currents due to the pair breaking process and the generation of quasiparticles. The dynamics can be probed as 1D signal of the peak TH-field changes $\delta E_{TH}$ as function of the pump-drive delay $\Delta t$ (signal along $\Delta t$, (c)) or as 2D signal of the transient THG at each time delay (Figs. 2 and 4). (b) The generated TH field without the optical pump pulse as a function of an internal THz gate delay time $t_g$ at the equilibrium superconducting state $T < T_c$. (c) The 1D signal of the photoinduced TH field changes $\delta E_{TH}$ as function of pump-drive time delay $\Delta t$ measured at the equilibrium peak field position indicated by the red arrow in (b). The onset of the signal shows pronounced oscillations at 2Ω and 4Ω of the driving THz field.

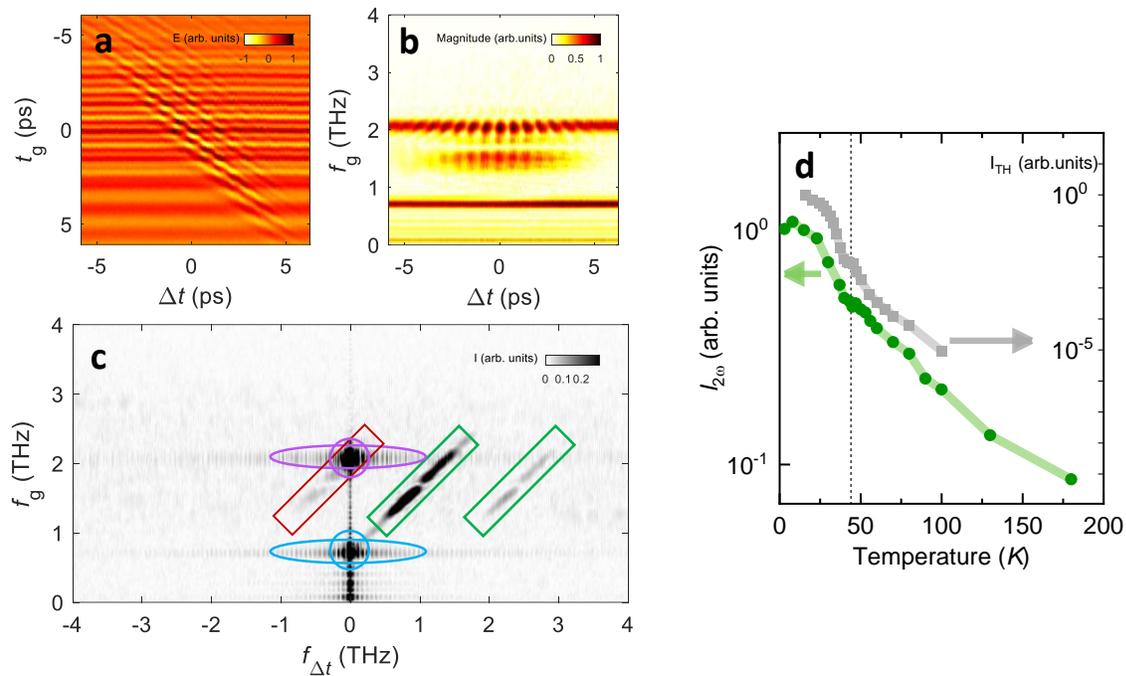

**Figure 2 2D signal of the nonlinear dynamics in the photoexcited LSCO.** (a) Transient dynamics of the transmitted THz field $E(t_g, \Delta t)$ at 5K after the excitation with an optical pump fluence $F_{pump} = 20 \mu J/cm^2$. The spectrum is normalized by the maximum value of the transient THz field. (b) Transient spectrum after a Fourier transformation of (a) along the THz gate time axis $t_g$. The spectrum is normalized by the equilibrium value of the FH magnitude. (c) 2D spectrum after transforming along gate time axis $t_g$ and pump-drive delay axis $\Delta t$. The circles mark components of the fundamental drive and the THG signal at equilibrium and the cigar-shaped marks represent the transient incoherent dynamics of the fundamental and TH field due to the optical pump pulse. The rectangular boxes mark a sideband generation that emerges from a modulation of the 3Ω TH signal with an Ω signal of quasiparticles in the driving THz field and higher orders, as seen by the oscillations at early times in the 1D scans of Fig. 1c. The discontinuous feature in the middle of the sideband is due to band-pass filters in the experimental setup (see Supplementary Note 3). (d) Intensity of the coherent 2Ω oscillation of the 3Ω TH signal ($I_{2\omega}$) compared to the static TH intensity $I_{TH}$ (Data from [25]) as function of temperature. The black dashed line marks the critical temperature $T_c$.

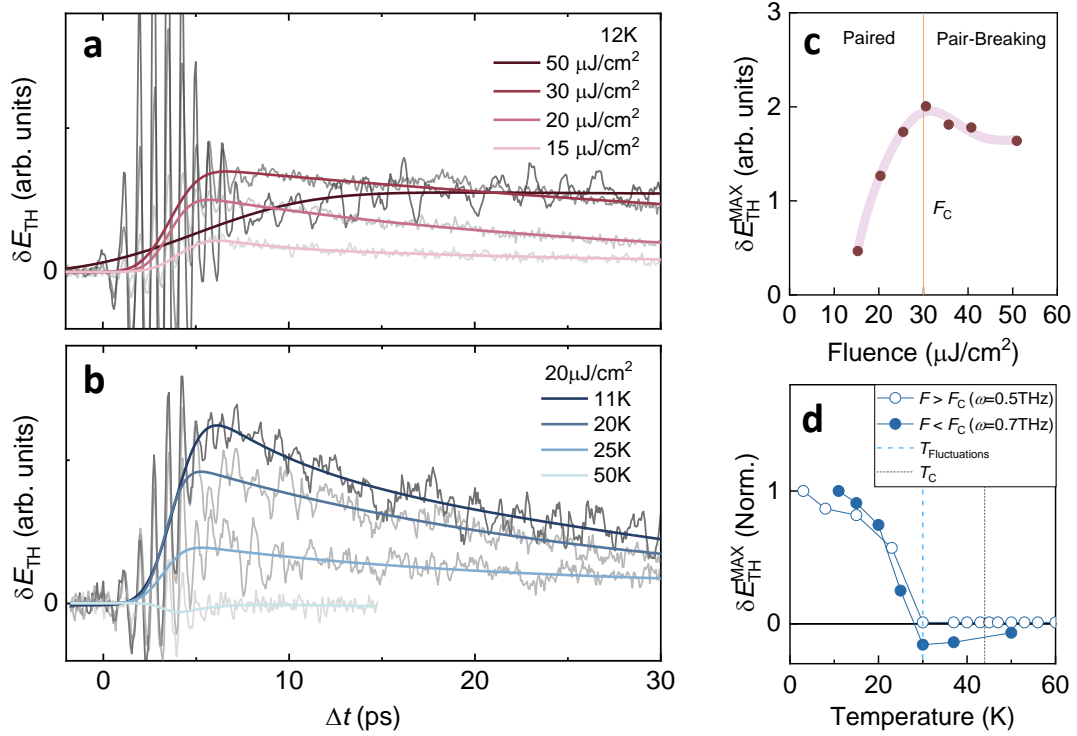

**Figure 3 Transient dynamics of the TH peak field amplitude in photoexcited LSCO.** (a) Pump fluence dependence at 12K. (b) Temperature dependence at a pump fluence $F_{pump} = 20 \mu J/cm^2$. (c) Fluence dependence of the extracted peak-field changes. The line shows a fit to the data with a square root increase and an additional linear decay that sets in at a critical pump fluence $F_c$ (red vertical line). The fit process is taken by ref. [42]. (d) Temperature dependence of the peak field changes at a pump flunces $F > F_C$ (open circles) and $F < F_C$ (filled circles). The blue dashed vertical line indicates the onset temperature for scaling fluctuation frequency as found in ref. 46. The black dotted vertical line indicates $T_c$.

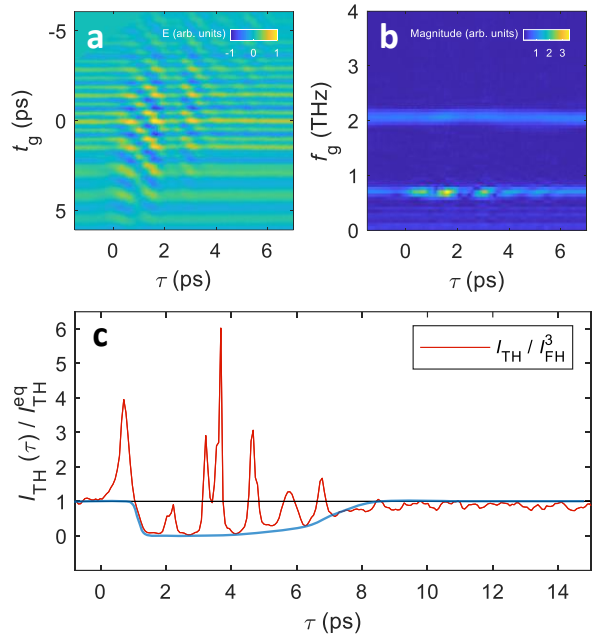

**Figure 4 Transient response of the 2D THG signal for the TH field at constant time delay .** (a) The data of Figure 2 (a) shearing in x axis ($\Delta t$), so that for each time delay $\tau = t_g + \Delta t$ the transient THz field has the same delay to the pump pulse. The spectrum is normalized by the maximum value of the transient THz field. (b) Fourier transform along the $t_g$ axis of (a). The spectrum is normalized by the equilibrium value of the FH magnitude. (c) Transient dynamics of the normalized TH susceptibility $I_{TH}/I_{FH}^3$. The value is again normalized by the dataset of the static case before the pump pulse arrives. The blue solid line is shown as a guide to the eye.

Supplementary information

# Tracing the dynamics of superconducting order via transient third harmonic generation

**Supplementary Note 1 Two methods for measuring the TH dynamics under the photoexcitation**

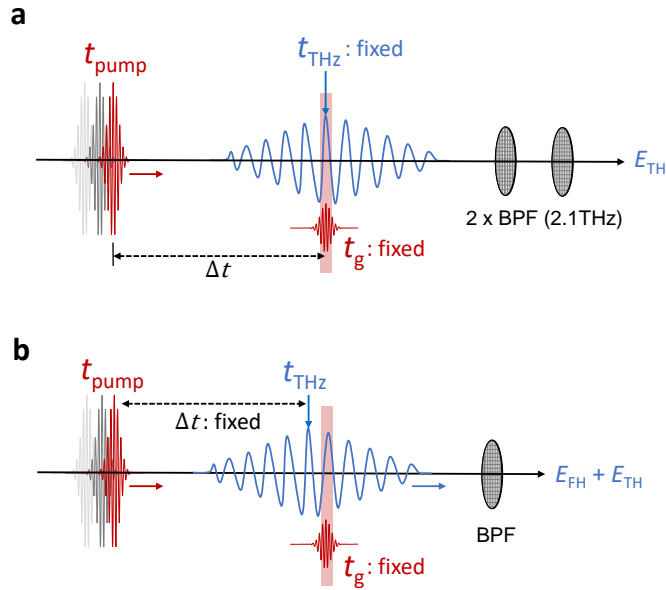

**Figure S1 Experimental scheme for the optical pump-TH drive experiment.** The three time points $t_{pump}$, $t_{THz}$ and $t_g$ are shown. $t=t_{THz}=0$ defines the center of the THz envelope and it is set conventionally to zero. The center $t=-t_{pump} <0$ of the pump pulse is located at negative times, and we denote the pump probe delays as $\Delta t = t_{THz} - t_{pump}$. In this convention, $\Delta t$ is always positive when the pump pulse precedes the THz probe. (a) 1D pump scheme. $t_{THz}$ and $t_{gate}$ are fixed to record the change of the TH field while sweeping $\Delta t$. To suppress the dominant FH field, two band pass filter (BPF) are installed after the sample. Here, $\Delta t$ changes in one scan. (b) 2D time-domain scans scheme. The $t_{pump}$ and $t_{THz}$ are simultaneously varying while the $t_g = 0$ is fixed. The $\Delta t$ is fixed during a scan in $t_{gate}$, subsequently each scan has a different value for the time distance $\Delta t$.

The Third harmonic (TH) field driven by high-field multi-cycle THz pulse is a new potential spectroscopic tool to investigate the superconducting order. By applying a combination of an optical pump pulse and a multicycle TH pulse, we investigate non-equilibrium superconductivity. Unlike the conventional time-domain THz probe spectroscopy, which represents the non-equilibrium linear electronic properties, the TH spectroscopy allows us to disentangle the nonlinear properties of the system from the pump-induced effects on the linear response, as due to quasiparticle excitations. We termed our novel protocol an *optical pump – TH drive* experiment, as opposed to a usual pump-probe scheme.

Tilted pulse front technique generates single-cycle broadband THz field. We filtered it to obtain a narrow-band driving field with the frequency centered at ω=0.7THz and average duration about 10 ps by using band-pass filter. To measure the driven TH electric field change under the photoexcitation, we focused the optical pump beam with the pulse duration of 100fs onto the sample with a beam diameter

600μm comparable to the focused THz pulse 560μm. The data acquisition processes were in two ways, 1) Scanning pump delay (1D scan) 2) Scanning pump and THz delay lines simultaneously (2D scan).

Figure S1 (a) depicts the 1D scan. Here, $t_{THz}$ and $t_g$ are kept fixed to record the change of the TH field while sweeping $\Delta t$. For this measurement it is necessary to fully suppress the leaked fundamental driving field (FH) field. We achieve this by installing two bandpass filters (2.1THz) after the sample. The optical pump beam is modulated by a chopper obtaining the relative changes of the TH field.

Figure S1 (b) shows the 2D scan scheme to collect the full THz spectrum. The static transmitted THz field is scanned by moving the $t_{THz}$ with the fixed gate time delay $t_g$. To collect the full THz spectrum including both FH and TH components, we used the one bandpass filter (2.1THz) after the sample. While the scanning, the optical pump pulse $t_{pump}$ moves with the fixed pump-drive time delay $\Delta t$ simultaneously. Each scans are operating with systematically varying $\Delta t$. In this procedure, we obtained the nonlinear sample response between the THz field and the optical pump pulse.

## Supplementary Note 2 Interference effect between optical pump and TH drive field

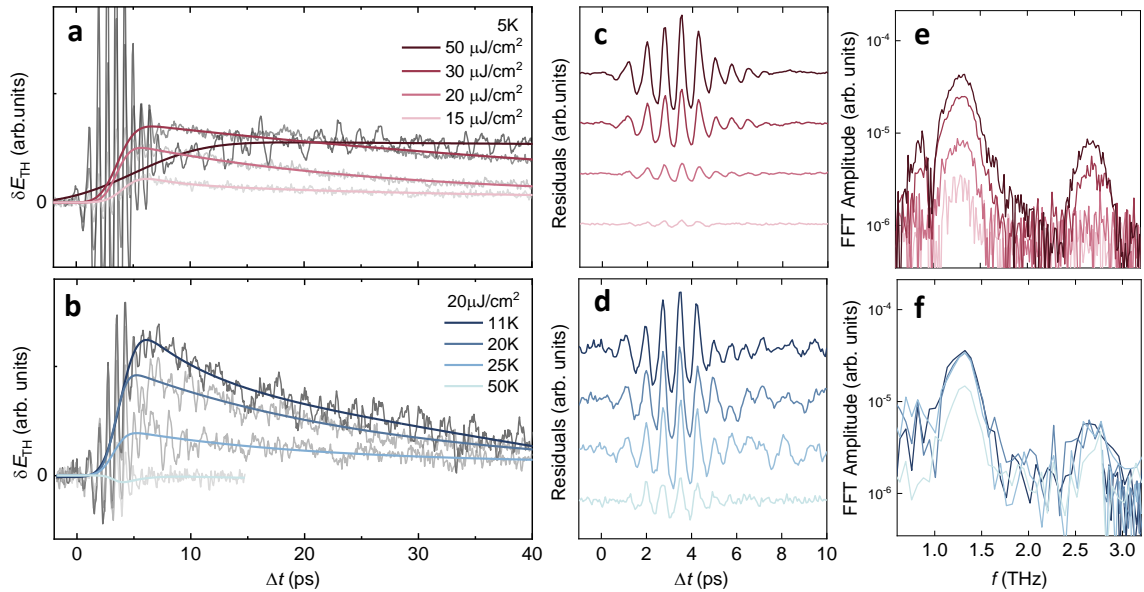

**Figure S2 Transient dynamics of the TH peak field under the photoexcitation.** (a,b) 1D pump-probe spectra of the experimentally measured photoexcited superconductivity probing nonlinearly driven TH signal in solid lines with greyscale. The fit curves are plotted with colored solid lines. (c,d) The coherent modulations are extracted by the fit of pairing amplitude dynamics. The FFT amplitude of the modulations are represented in (e) and (f).

Figure S2 describes the way of extracting the photoexcited interference feature in a superconductor. At the early pump-drive time delay $\Delta t$ a prominent strong oscillatory signal emerges on top of the background response, as seen in Fig. S2 (a,b). To extract the oscillatory signal, we fit the non-oscillatory background signal in analogy to previous studies on other correlated materials [1-3]. The fit function is designed as

$$\delta E_{TH} = \left(1 + \text{erf}\left(\frac{\Delta t - \Delta t_0}{\sqrt{2}\tau_{el}}\right)\right)\left(A_1 \exp\left(-\frac{\Delta t - \Delta t_0}{\tau_1}\right) + A_2 \exp\left(-\frac{t - \Delta t_0}{\tau_2}\right)\right) + c, \quad (S1)$$

where $\Delta t$ is the time delay, $\tau_{el}$ is the time constant for the excitation of photo-carriers, $\tau_1$ and $\tau_1$ are the time constant of the exponential decay, and $A_1$, $A_2$, $c$, $\Delta t_0$ are constants. Subtracting the fits results in Fig. S2 (c,d). As explained at the end of the previous section, in the 1D pump scan we select only the $\omega_g = 3\Omega$ dynamics of $\delta E_{TH}$ by adding an additional band-pass filter. As a consequence the frequency of the $\Delta t$ oscillations are $\omega_{\Delta t} = \pm 2\Omega$ (1.4THz) and $\omega_{\Delta t} = \pm 4\Omega$ (2.8THz), see Fig. S2 (e,f). Apart from the (non-oscillatory) contribution of the static THG signals, Fig. S2 (c,d) are equivalent to the projection of the 2D scans with digitally filtered to obtain only TH component as described in the Supplementary Note 3.

### Supplementary Note 3 Theoretical calculation of pump-TH drive dynamics

To pinpoint the origin of THG modulations with even multiples of $\Omega$ as reported in the main text, we recorded spectra of the transmitted signal as a function of gate time $t_g$ and pump-drive delay $\Delta t$. The resulting spectrum is shown in Fig. S3 (a). Fig. S3 (b-c) show the data after a Fourier transform first along the $t_g$ axis, and then along the $\Delta t$-axis, with corresponding frequency variables $\omega_g$ and $\omega_{\Delta t}$, respectively. In what follows, we will first focus on the discussion of the experimental data in $(\omega_g, \omega_{\Delta t})$-space shown in Fig. S3 (c). The experimentally recorded transmitted field is directly related to the induced current $j(\omega) = j^{(1)}(\omega) + j^{(3)}(\omega)$ which we further decompose into first- and third-order contributions of the applied electromagnetic vector potential $A = A_{THz} + A_{pump}$. Signatures of the fifth-order current $j^{(5)}(\omega)$ are also observed, as detailed in the below.

The experimental setup consists of three pulses. The optical pump $A_{pump}$, the THz drive $A_{THz}$, and a gate pulse at which the parallel electric field component of the transmitted beam is recorded. Here we define the time origin $t = t_{THz} = 0$ to be the center of the THz drive. We denote functional forms of the pulse shapes centered around the origin by a bar, so that $A_{pump}(t) = \bar{A}_{pump}(t + \Delta t)$. In frequency space the pump delay $\Delta t$ hence results in a phase shift $\hat{A}_{pump}(\omega) = \bar{A}_{pump}(\omega) e^{-i\omega \Delta t}$. The linear response current is given by

$$j^{(1)}(\omega_g) = T(\omega_g)\chi^{(1)}(\omega_g)A(\omega_g) \approx \chi^{(1)}(\omega_g)A_{THz}(\omega_g), \quad (S2)$$

where $\chi^{(1)}$ is the linear susceptibility and $T(\omega)$ is the transmission function of a $3\Omega$-bandpass filter shown in Fig. S4. Only the probe pulse generates a response in the THz regime. Since $A_{THz}$ is unaffected by $\Delta t$ and is a narrowband multi-cycle pulse centered at frequency $\Omega$, the linear response yields the signal at $(\omega_g, \omega_{\Delta t}) = (\pm \Omega, 0)$ in Fig. S3 (c).

The third order current response is given by the general expression [4]

$$j^{(3)}(\omega_g) = T(\omega_g)\int d\omega_1 d\omega_2 \chi^{(3)}(\omega_g, \omega_1, \omega_2)A(\omega_1)A(\omega_2)A(\omega_g - \omega_1 - \omega_2), \quad (S3)$$

where $\chi^{(3)}$ is the nonlinear susceptibility that captures all microscopic details of the sample, such as the quasiparticles and the Higgs contributions. Decomposing the vector potential as $A = A_{THz} + A_{pump}$, we obtain four different contributions (Fig. S5). The terms $j^{(3)} \sim A_{THz}^2 A_{pump}$ (Fig. S5 (a)) and $j^{(3)} \sim A_{pump}^3$ (Fig. S5 (b)) are odd in orders of the pump pulse and vanish in the THz regime. $j^{(3)} \sim A_{THz}^3$ (Fig. S5 (c)) is

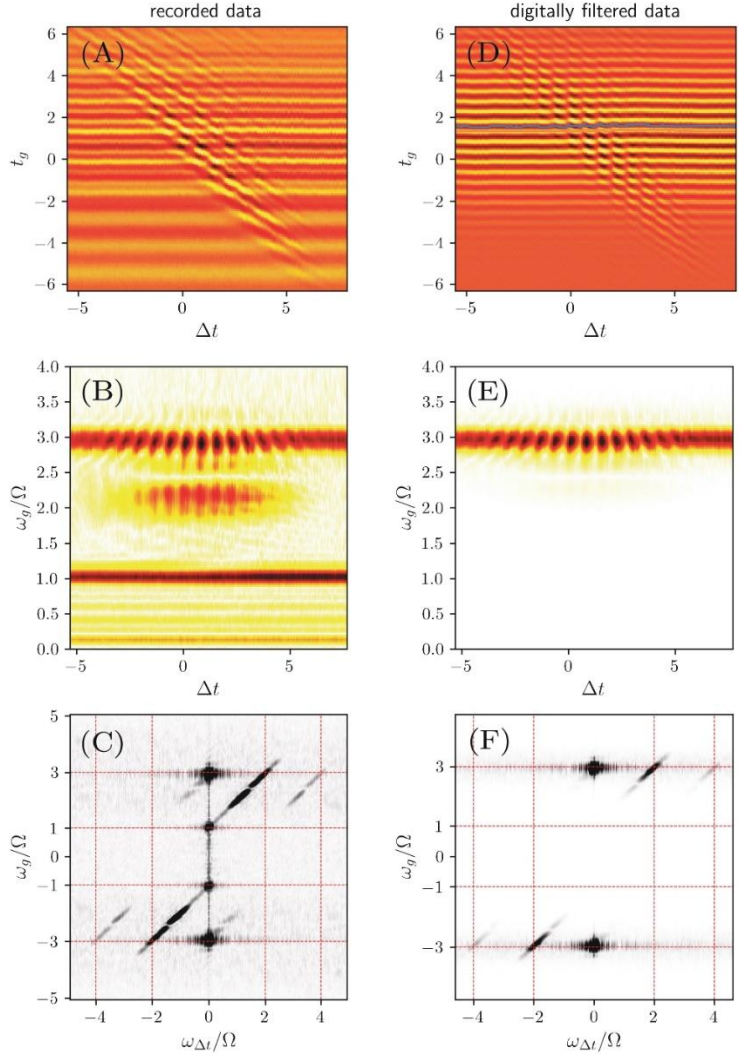

**Figure S3** (a-c) Experimentally recorded data for 2D scans in (a) time domain, (b) time-frequency domain and (c) frequency domain. The experiment was performed using a single 2.1 THz band pass filter, so that leakage of FH can be observed. (d-f) Projection of the 2D scan results on the experimental conditions used to acquire the 1D scans. Since in 1D scans two band-pass 2.1 Thz filters have been used, a second 2.1 THz band pass filter has been digitally applied to the data of panel (a-c) to match the experimental setup used to produce the data shown in Fig. 1(c) of the main manuscript and in Supplementary Note 2. This completely filters out the FH signal, as seen in panels (e) and (f). In addition, the second filter also removes the TH modulations at $\omega_{\Delta t} = 4\Omega$ that are still visible in panel (f). The data Fourier transformed back to real time are shown in panels (d) and (e).

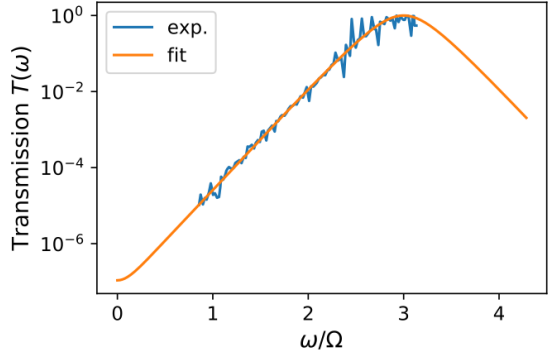

**Figure S4** Transmission function $T(\omega)$ of the filter. Blue curve shows the experimentally recorded data, orange curve shows a fit used in the data processing routine.

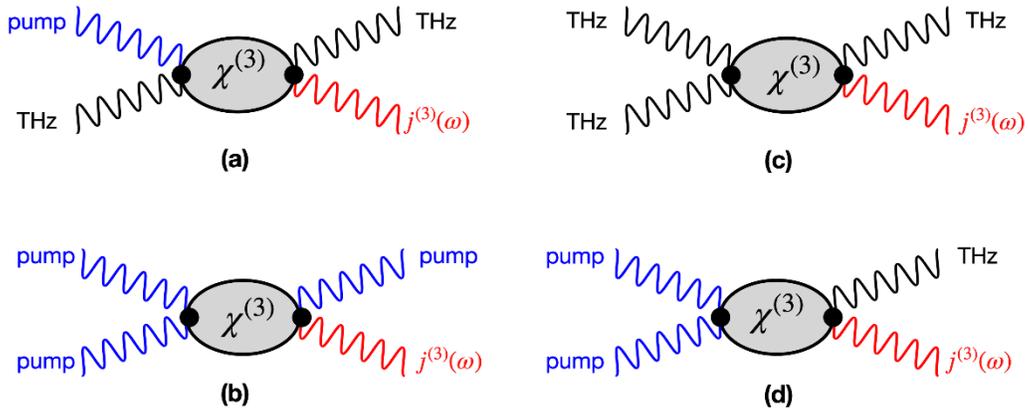

**Figure S5** Diagrammatic representation of all the contributions to the third-order nonlinear current $j^{(3)}(\omega)$ (red wiggly line). The third-order susceptibility $\chi^{(3)}$, including in principle all diamagnetic, paramagnetic and mixed terms in the presence of disorder (see Ref. [4] for more details), is schematically represented here as a full grey bubble. (a) Process involving two THz and one optical pump photon: the nonlinear current response in this case is in the optical range as the pump field. (b) FH and TH generation process of the optical pump field. (c) FH and TH processes of the THz field, which provides the contributions to the $\Omega$ and $3\Omega$ signals measured in the experiments. (d) Process involving two pump photons (blue) and one THz photon (black): in order to obtain a THz response a difference-frequency process involving the two pump photons needs to be considered.

the usual pump-independent THG response that gives the $(\omega_g, \omega_{\Delta t}) = (\pm 3\Omega, 0)$ signal in Fig. S3 (c), as well as an additional contribution at $(\omega_g, \omega_{\Delta t}) = (\pm \Omega, 0)$ [4]. The last term $j^{(3)} \sim A_{THz} A^2_{pump}$ (Fig. S5 (d)) deserves a more detailed discussion. Approximating the narrowband THz probe as $A_{THz}(\omega) = A^0_{THz}(\omega)\delta(\omega \pm \Omega)$, we obtain from Eq. (S3)

$$j^{(3)}(\omega_g, \Delta t) = T(\omega_g) \int d\omega_2 \chi^{(3)}(\omega_g, \pm\Omega, \omega_2) A^0_{THz}(\pm\Omega) \hat{\bar{A}}_{pump}(\omega_2) \hat{\bar{A}}_{pump}(\omega_g \mp \Omega - \omega_2) e^{-i(\omega_g \mp \Omega)\Delta t}.$$
(S4)

The $\Delta t$-dependence of the nonlinear current is solely manifest in the $e^{-i(\omega_g \mp \Omega)\Delta t}$ phase. The Fourier component of the current at frequency $\omega_g$ hence shows phase oscillations in $\Delta t$ with frequency $\omega_g \pm \Omega$. Once the signal is further Fourier transformed with respect to $\Delta t$ one readily obtains from Eq. (S4) the continuous signal along the diagonal lines at $\omega_{\Delta t} = \omega_g \pm \Omega$ in Fig. S3 (c). The width of the signal is limited to the regime $(A_{pump} \otimes A_{pump})(\omega_g \pm \Omega)$ in the $\omega_g$-direction, where $\otimes$ denotes a convolution.

For the optical pump pulse used in the experiment, $A_{pump} \otimes A_{pump}$ has an approximately Gaussian peak centered around zero with FWHM of ~5 THz, consistent with the broad diagonal feature observed between $|\omega_g| \lessapprox 3\Omega$. The absence of signal at low frequencies (between $|\omega_g| \lessapprox \Omega$) in Fig. S3 (f) is a consequence of the $3\Omega$ bandpass filter used in the experiment. We can intuitively understand Eq. (S4) within the diagrammatic representation shown in Fig. S5. Here, one photon of the THz drive pulse (black wavy line) combines with two photons of the pump pulse (blue wavy lines) to produce the nonlinear current (red wavy line). The full grey bubble represents the nonlinear susceptibility $\chi^{(3)}$ that embodies the microscopic details of the material. In the case of both the quasiparticle and the Higgs mode contributions, such renormalization would lead to a peak in the susceptibility at the characteristic resonance energy of $2\Delta$. To conserve energy, the frequency of the three photons must match the frequency of the induced current. The probe photon can only contribute with energies $\pm\Omega$. While the photon energies of the pump pulse lie in the optical regime, i.e. $\nu \sim \pm eV$, their frequency spectrum is rather broad and two such high energy photons can combine in a difference frequency (DF) process to yield the total energy $\omega \mp \Omega$ in the THz regime. Due to the pump delay $\Delta t$, both pump photons pick up a phase that results in phase oscillations in $\Delta t$ manifest in the diagonal spectral features at $\omega_{\Delta t} = \omega_g \pm \Omega$ in Fig. S3 (c) (marked by the green rectangles boxes in Fig. 2 (c) in the main text). Following the same steps detailed above, the diagonal stripes at $\omega_{\Delta t} = \omega_g \pm 3\Omega$ in Fig. S3 (c) (marked by the red rectangle box in Fig. 2 (c) in the main text) can be related instead to the fifth-order current $j^{(5)}$.

The 2D experiment described above differs from the 1D experiment described in the main text in two ways. First, the 1D experiment was performed at two selected values of the gate time $t_g = t^\pm$, corresponding to the maximum(minimum) field strength of the THz probe pulse. Second, the 2D experiment was performed using a single $3\Omega$ bandpass filter of transmission $T(\omega)$, see Eq. (S2), while two such filters, $T(\omega)^2$, were used in the 1D experiment. The experimentally recorded transmission function of the filter is shown in Fig. S4. To verify the consistency of the two experiments, we bridge these difference in two steps. We first digitally apply the second filter to the 2D data according to

$$j^{(3)}_{filtered}(\omega_g) = T(\omega_g) j^{(3)}(\omega_g) \text{ (S5)}$$

The filtered data in the $(\omega_g, \omega_{\Delta t})$-domain are shown in Fig. S3 (f). We see that the remaining signal is concentrated in a narrow band around $\omega_g = \pm 3\Omega$. Fourier transforming the axes back into time domain yields Fig. S3 (d-e).

Next, we tune the gate time $t_g$ to the maximum (minimum) $t^\pm$ of the signal as indicated by blue (orange) lines in Fig. S3 (d). In principle $t^\pm$ should be independent of $\Delta t$, but we allow for a small $\Delta t$-dependence to compensate for small drifts of the pulse position when $\Delta t$ is experimentally scanned.

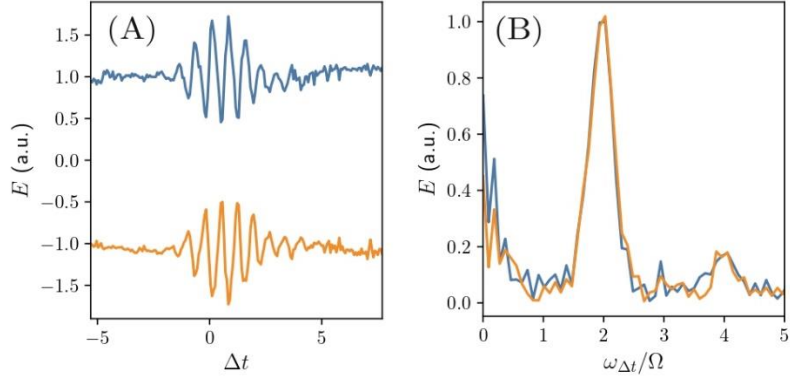

**Figure S6** (a) Plot of the maximum (blue) and minimum (orange) of the digitally filtered data of Fig. S3 (d) as a function of $\Delta t$. The data correspond to the slice marked by horizontal lines in Fig. S3 (d). The Fourier transform of the data with respect to $\Delta t$ in (b) reveals $2\Omega$ and $4\Omega$ components of the modulations in $\Delta t$.

A plot of the signal along the blue (orange) lines is shown in Fig. S6 (a). We observe oscillations in $\Delta t$ in agreement to the 1D data of Fig. S2 (a,b) (Fig. 3 (a,b) in the main text). The corresponding Fourier transform in Fig. S6 (b) reveals the frequency of these oscillations to be $2\Omega$ and $4\Omega$.

From the preceding discussion, we can understand the origin of the $2\Omega$ and $4\Omega$ oscillations. The measurement for a specific gate time $t^{\pm}$ corresponds to an integration over a range of $\omega_g$ frequencies in Fourier space. This is approximately equivalent to a projection of the spectrum in Fig. S3 (c) on the $\omega_{\Delta t}$-axis. However, the presence of the filter removes all but the $\omega_g = \pm 3\Omega$ signal. Hence, mostly the signal at $\omega_g = \pm 3\Omega$ contributes. Therefore, the 1D measurement effectively probes the points $(\omega_g, \omega_{\Delta t}) = (\pm 3\Omega, 0)$, $(\pm 3\Omega, \pm 2\Omega)$, and $(\pm 3\Omega, \pm 4\Omega)$ in the 2D FT spectrum shown in Fig. S3 (f). The same effect is easily seen in time domain in the 1D differential scans reported in Fig. 1 (c) and Fig. 3 (a,b) of the main text. In this case the THz field-change $\delta E_{TH}$ scales as the amplitude of the pump-induced contribution $j^{(3)}(\omega_g, \Delta t)$ to the non-linear current given by Eq. (S4). It is then expected that it oscillates as a function of $\Delta t$ at $2\Omega$ and $4\Omega$, the latter being suppressed by the filtering procedure as shown by comparing Fig. S4 (c) and Fig. S4 (f).

**Supplementary Note 4 Comparison the pump induced TH changes to unpumped TH amplitude**

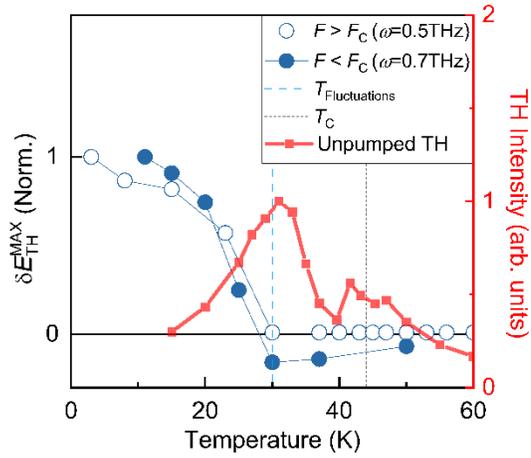

**Figure S7 Temperature dependence of the TH peak field changes and unpumped TH field intensity.** The blue data (left axis) shows the TH peak field changes under the photoexcitation as discussed in the main text (Fig. 4 (d)) and the red data (right axis) shows the TH intensity without optical pump from Ref. [5].

Figure S7 shows the direct comparison of the TH field changes under the photoexcitation to the unpumped TH field response. The onset of the TH field changes signal below 30K appears at temperatures where the unpumped TH field intensity peaks. As discussed in ref. 5 this suppression of the static TH signal is due to strong strong screening effects that take place in the superconducting phase [5]. This finding suggests that the sign-change of the pump- induced variation of the TH signal can be explained by a predominant effect of the local-field screening.

## Supplementary Note 5 2D pump-drive spectra along sheared time axis $\tau$

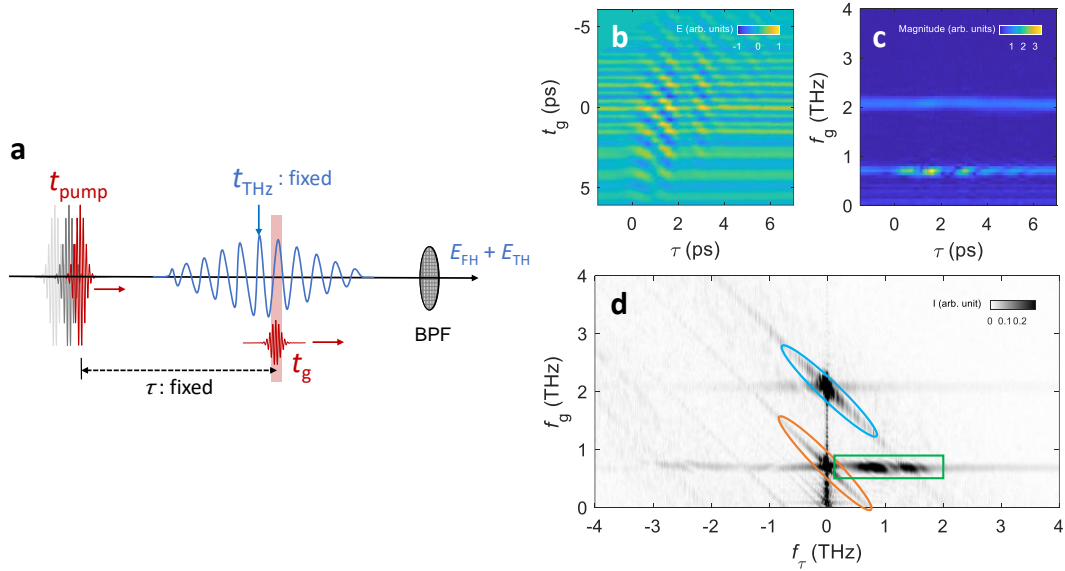

**Figure S8 Intrinsic response of optical pump-drive signal.** (a) Equivalent experimental scheme to shearing 2D spectrum. The $t_{pump}$ and $t_{gate}$ are simultaneously varying while the $t_{THz}$ =0 is fixed. The time difference between the pump and the gate, $\tau$, is fixed during a scan, subsequently each scan has a different value for the time distance $\tau = t_g + \Delta t$. (b) The sheared data in x-axis ($\Delta t$) of Figure 2(a). The spectrum is normalized by the maximum value of the transient THz field. (c) Fourier transform along the $t_g$ axis of (a). The spectrum is normalized by the equilibrium value of the FH magnitude. (d) A 2D Fourier transform of (b) along the sampling $t_g$ and pump-drive delay axis $\tau$. Cigar- and rectangular- shaped features correspond to pumped dynamics and emerged photoexcited harmonics, respectively.

To examine the features of the transient dynamics of the linear FH component and the nonlinear TH signal distinctly, we now consider a different scheme: namely, we introduce a new variable $\tau = t_g + \Delta t$ and plot the 2D spectra as a function of $(t_g, \tau)$ instead of $(t_g, \Delta t)$, as shown in Fig. S8 (b). Geometrically, this corresponds to a 45° shear of the original 2D spectrum in Fig. S3 (a). This physically corresponds to collecting individually traces of the experimental data at fixed pump-gate delay, or adopting the method 2 described in Ref. [6]. This is equivalent to the experimental scheme shown in Fig. S8 (a). By fixing the time between the pump pulse and gate pulse and sweeping both pulses while the THz pulse is fixed, every point of the THz waveform is measured with same pump-probe delay time $\tau$.

As shown in Fig. S8 (c) after the Fourier transform along the gate time $t_g$, the FH now shows coherent modulations as a function of pump-drive delay time $\tau$. The spectral feature of background excitation-relaxation dynamics are marked by cigar-shaped contours (blue: TH, orange: FH). These features are tilted by 45° compared to the corresponding in the $(f_g, f_{\Delta t})$-representation of (Fig. 2 (c)). In a similar vein, the horizontal feature at the FH component shown in frequency with broadened from ω to 2ω marked by green rectangle in Fig. S8 (d) indicates that the FH modulations is highly related on the TH modulations.

# Experimental Method

**Sample preparation**

The optimally doped LSCO sample was grown by molecular beam epitaxy (MBE) at the Max Planck Institute for Solid State Research. The sample is 80 nm-thick on a LaSrAlO$_4$ (LSAO) substrate. $T_c$ was determined at 45K from mutual inductance measurement.

**Experiment**

We performed an optical pump and THG drive experiment with THz sources based on a femtosecond laser system. For the TH generation, broadband THz radiation was generated through tilted pulse front scheme utilizing lithium niobate crystal [1-3]. With initial laser pulse energy around 1.5 mJ at 800 nm central wavelength and 100 fs pulse duration broadband THz radiation with up to 3 µJ pulse energy was generated. To produce narrow-band THz radiation, corresponding bandpass filters were used; the set-up of TH generation scheme in detail in ref. [4]. In the photoexcitation measurement with a NIR pump pulse, an 1D pump-TH drive scheme shown in Fig. 1c and Fig. 3 were performed by varying of the time delay between the pump pulse and the gate pulse by moving the pump pulse. On the other hand, a 2D pump-TH drive experiment was performed by sweeping the THz pulse with a fixed time delay between the THz pulse and the NIR pump pulse with a fixed gate pulse. See Supplementary Note 1 for more information.